\documentclass[twoside]{dis04}
\def\runauthor{M. Diehl,\, B.~Pire,\, L.~Szymanowski}
\def\shorttitle{DEEP ELECTROPRODUCTION OF PENTAQUARKS}
\begin{document}
\newcommand{\lsim}{\mathrel{\vcenter
    {\hbox{$<$}\nointerlineskip\hbox{$\sim$}}}}
\newcommand{\gsim}{\mathrel{\vcenter
    {\hbox{$>$}\nointerlineskip\hbox{$\sim$}}}}

\newcommand{\re}{\mathrm{Re}\,}
\newcommand{\im}{\mathrm{Im}\,}

\newcommand{\gev}{\,{\rm GeV}}
\newcommand{\mev}{\,{\rm MeV}}

\newcommand{\half}{\textstyle \frac{1}{2}}

\newcommand{\mth}{m_\Theta}

\newcommand{\eth}{\eta_\Theta}

\title{EXCLUSIVE ELECTROPRODUCTION OF PENTAQUARKS
}

\author{M. Diehl$^{a}$,\, B.~Pire$^b$,\, L.~Szymanowski$^c$}

\address{${}^a$\,Deutsches Elektronen-Synchroton DESY, 22603 Hamburg, Germany
\\
${}^b$\,CPHT, {\'E}cole Polytechnique, 91128 Palaiseau, France
\\
${}^c$\,Soltan Institute for Nuclear Studies, Ho\.{z}a 69, 00-681
Warsaw, Poland \\ }

\maketitle

\abstracts{Exclusive electroproduction of a $K$ or $K^*$
meson on the nucleon can give a $\Theta^+$ pentaquark in the final
state.  This reaction offers an opportunity to investigate the
structure of pentaquark baryons at parton level.  We discuss the
generalized parton distributions for the $N \to \Theta^+$ transition
and give the leading order amplitude for electroproduction in the
Bjorken regime.  
}

\section{Introduction}

There is increasing experimental evidence
\cite{Nakano:2003qx,Barth:2003es} for the existence of a narrow baryon
resonance $\Theta^+$ with strangeness $S=+1$, whose minimal quark
content is $uudd\bar{s}$.  Triggered by the prediction of its mass and
width in \cite{Diakonov:1997mm,Praszalowicz:2003ik}, the observation 
of this hadron
promises to shed new light on our picture of baryons in QCD.  
A fundamental question is
 how the structure of baryons manifests itself in terms of the basic
degrees of freedom in QCD, at the level of partons.  This structure at
short distances can be probed in hard exclusive scattering processes
\cite{browder},
where it is encoded in generalized parton distributions
 \cite{Muller:1994fv} (see \cite{Diehl:2003ny} for a recent
review).  In Ref. \cite{DPS} we introduced the generalized parton 
distributions (GPDs) for the transition from the
nucleon to the $\Theta^+$ (denoted as $\Theta$ below) and investigated
electroproduction processes
where these GPDs could be measured, hopefully already in existing
experiments at DESY and Jefferson Lab.


\section{Processes}
\label{sec:channels}

We consider the electroproduction processes
\begin{equation}
  \label{proc-p}
e p\to e \bar{K}^0 \, \Theta  , \qquad \qquad
e p\to e \bar{K}^{*0} \, \Theta ,
\end{equation}
where the $\Theta$ subsequently decays into $K^0 p$ or $K^+ n$.  Note
that the decay $\bar{K}^{*0} \to K^- \pi^+$ of the ${K}^{*}(892)$ tags
the strangeness of the produced baryon.  In contrast, the observation
of a $\bar{K}^0$ as $K_S$ or $K_L$ includes a background from final
states with a $K^0$ and an excited $\Sigma^+$ state in the mass region
of the $\Theta$, unless the strangeness of the baryon is tagged by the
kaon in the decay mode $\Theta\to K^+ n$.  Apart from their different
experimental aspects the channels with $\bar{K}$ or $\bar{K}^*$
production are quite distinct in their dynamics.  
The crossed process $K^+ n\to e^+e^-\,
\Theta$ could be analyzed along the lines of \cite{Berger:2001zn} at
an intense kaon beam facility.

\begin{figure}
\begin{center}
\leavevmode
\psfig{figure=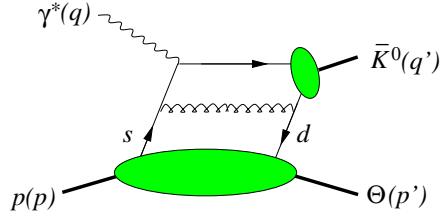,width=0.45\textwidth}
\end{center}
\caption{\label{fig:meson} One of the graphs for the $\gamma^* p \to
\bar{K}^0 \, \Theta$ amplitude in the Bjorken limit.  The large blob
denotes the GPD for the $p\to \Theta$ transition and the small one the
DA of the kaon.}
\end{figure}

The kinematics of the $\gamma^* p$  subprocess is
specified by the invariants
$¥Q^2 = - q^2 ,
W^2 = (p+q)^2 , 
t = (p-p')^2 $,
with four-momenta as given in Fig.~\ref{fig:meson}.  We are interested
in the Bjorken limit of large $Q^2$ at fixed $t$ and 
$x_B = Q^2 /(2 pq)$.

According to the factorization theorem for meson production
\cite{Collins:1997fb}, the Bjorken limit implies factorization of the
$\gamma^* p$ amplitude into a perturbatively calculable subprocess at
quark level, the distribution amplitude (DA) of the produced meson,
and a generalized parton distribution (GPD) describing the transition
from $p$ to $\Theta$ (see Fig.~\ref{fig:meson}).  The dominant
polarization of the photon and (if applicable) the produced meson is
then longitudinal, and the corresponding $\gamma^* p$ cross section
scales like $d\sigma_L /(dt) \sim Q^{-6}$ at fixed $x_B$ and $t$, up
to logarithmic corrections in $Q^2$ due to perturbative evolution.


\section{The transition GPDs and their physics}
\label{sec:gpds}

  To define the transition GPDs we introduce
light-cone coordinates $v^\pm = (v^0 \pm v^3) /\sqrt{2}$ and
transverse components $v_T = (v^1, v^2)$ for any four-vector $v$.  The
skewedness variable $\xi = (p-p')^+ /(p+p')^+$ describes the loss of
plus-momentum of the incident nucleon and is connected with $x_B$ by
$\xi \approx {x_B}/{(2-x_B)}$
in the Bjorken limit.

In the following we assume that the $\Theta$ , which we treat as a
stable hadron, has spin $J=\half$ and
isospin $I=0$.  Different theoretical approaches predict either $\eth
= 1$ or $\eth = -1$ for the intrinsic parity of the $\Theta$, and we
will give our discussion for the two cases in parallel \cite{Rekalo}. 
 The
hadronic
matrix elements that occur in the electroproduction processes
(\ref{proc-p}) at leading-twist accuracy are
\begin{eqnarray}
  \label{matrix-elements}
F_V &=&
\frac{1}{2} \int \frac{d z^-}{2\pi}\, e^{ix P^+ z^-}
  \langle \Theta|\, \bar{d}(-\half z)\, \gamma^+ s(\half z) 
  \,|p \rangle \Big|_{z^+=0,\, {z}_T=0} \; ,
\nonumber \\
F_A &=&
\frac{1}{2} \int \frac{d z^-}{2\pi}\, e^{ix P^+ z^-}
  \langle \Theta|\, 
     \bar{d}(-\half z)\, \gamma^+ \gamma_5\, s(\half z)
  \,|p \rangle \Big|_{z^+=0,\, {z}_T=0}\;,
\end{eqnarray}
with $P = \half (p+p')$, where here and in the following we do not
explicitly label the hadron spin degrees of freedom.  We define the
corresponding $p\to \Theta$ transition GPDs by
\begin{eqnarray}
  \label{gpd-pos}
F_V &=& \frac{1}{2P^+} \left[
  H(x,\xi,t)\, \bar{u}(p') \gamma^+ u(p) +
  E(x,\xi,t)\, \bar{u}(p') 
                 \frac{i \sigma^{+\alpha} (p'-p)_\alpha}{\mth+m_N} u(p)
  \, \right] ,
\nonumber \\
F_A &=& \frac{1}{2P^+} \left[
  \tilde{H}(x,\xi,t)\, \bar{u}(p') \gamma^+ \gamma_5 u(p) +
  \tilde{E}(x,\xi,t)\, \bar{u}(p') 
\frac{\gamma_5\, (p'-p)^+}{\mth+m_N} u(p)
  \, \right]
\end{eqnarray}
for $\eth = 1$ and by
\begin{eqnarray}
  \label{gpd-neg}
F_V &=& \frac{1}{2P^+} \left[
  \tilde{H}(x,\xi,t)\, \bar{u}(p') \gamma^+ \gamma_5 u(p) +
  \tilde{E}(x,\xi,t)\, \bar{u}(p') 
\frac{\gamma_5\, (p'-p)^+}{\mth+m_N} u(p)
  \, \right] ,
\nonumber \\
F_A &=& \frac{1}{2P^+} \left[
  H(x,\xi,t)\, \bar{u}(p') \gamma^+ u(p) +
  E(x,\xi,t)\, \bar{u}(p') 
        \frac{i \sigma^{+\alpha} (p'-p)_\alpha}{\mth+m_N} u(p)
  \, \right]
\end{eqnarray}
for $\eth = -1$.  
The scale dependence of the matrix elements is governed by the
nonsinglet evolution equations for GPDs
\cite{Muller:1994fv,Blumlein:1997pi}, with the unpolarized evolution
kernels for $F_V$ and the polarized ones for $F_A$. 

\begin{figure}
\begin{center}
\leavevmode
\psfig{figure=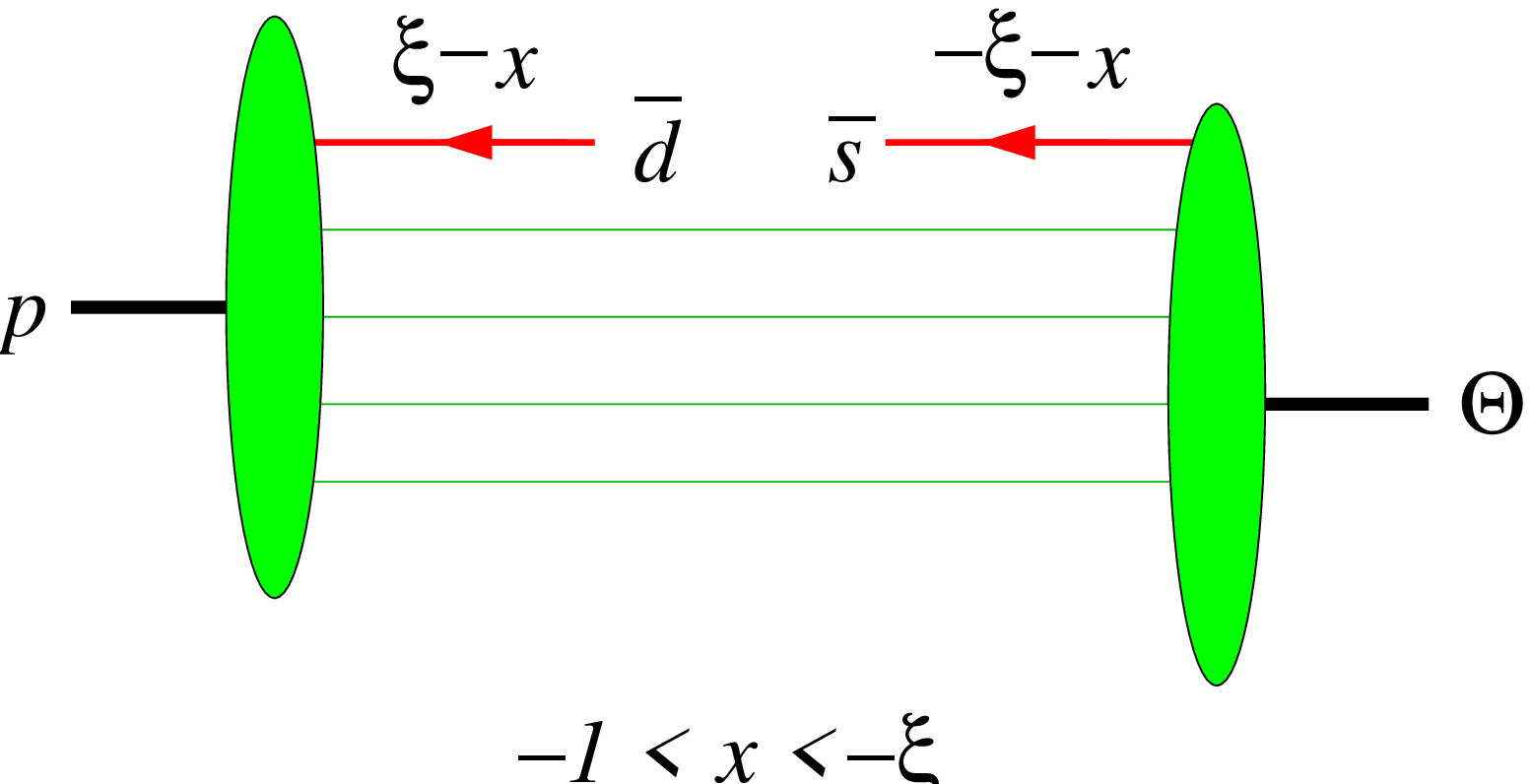,width=0.32\textwidth}
\hspace{1em}
\psfig{figure=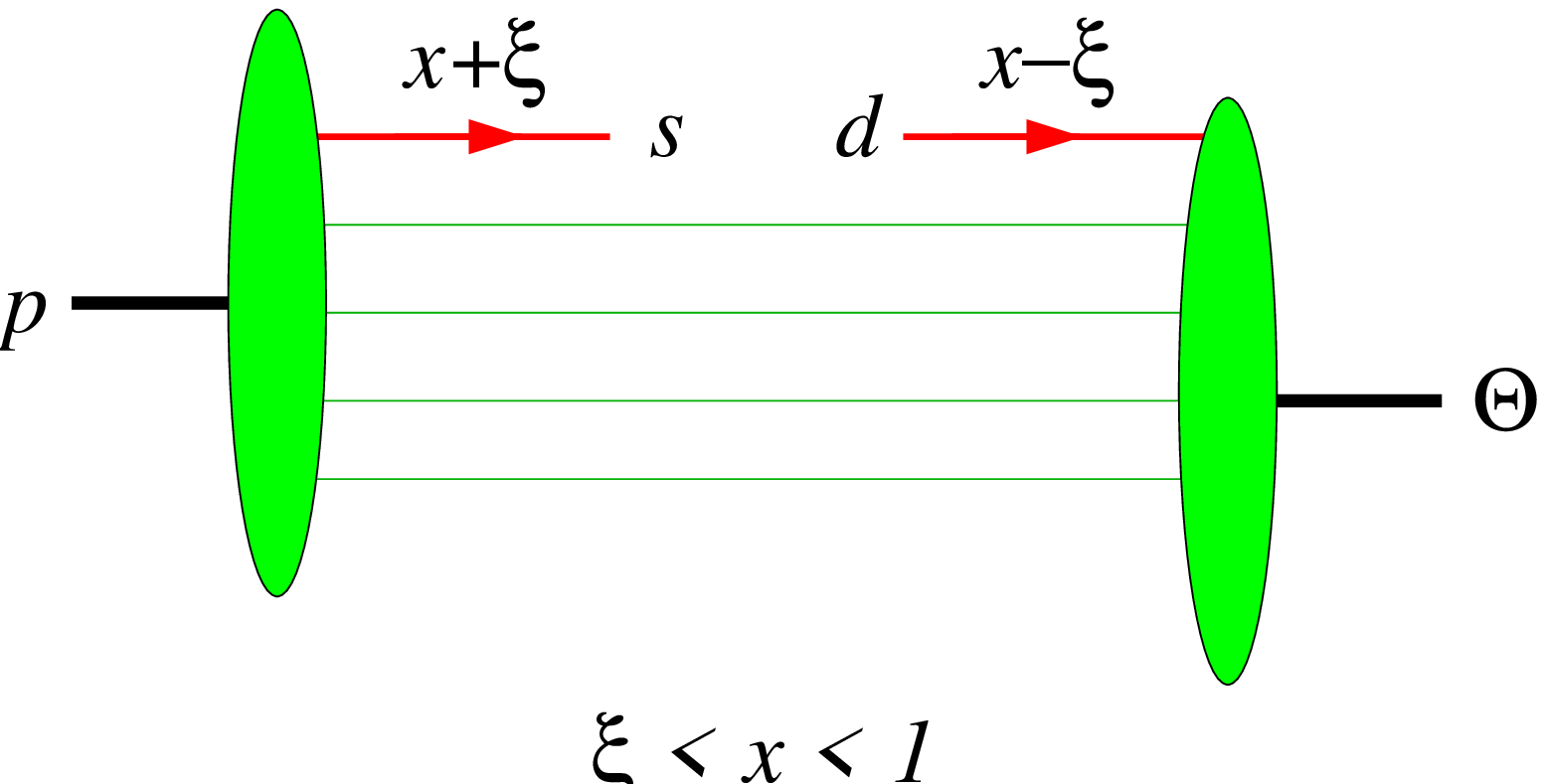,width=0.32\textwidth}
\hspace{1em}
\psfig{figure=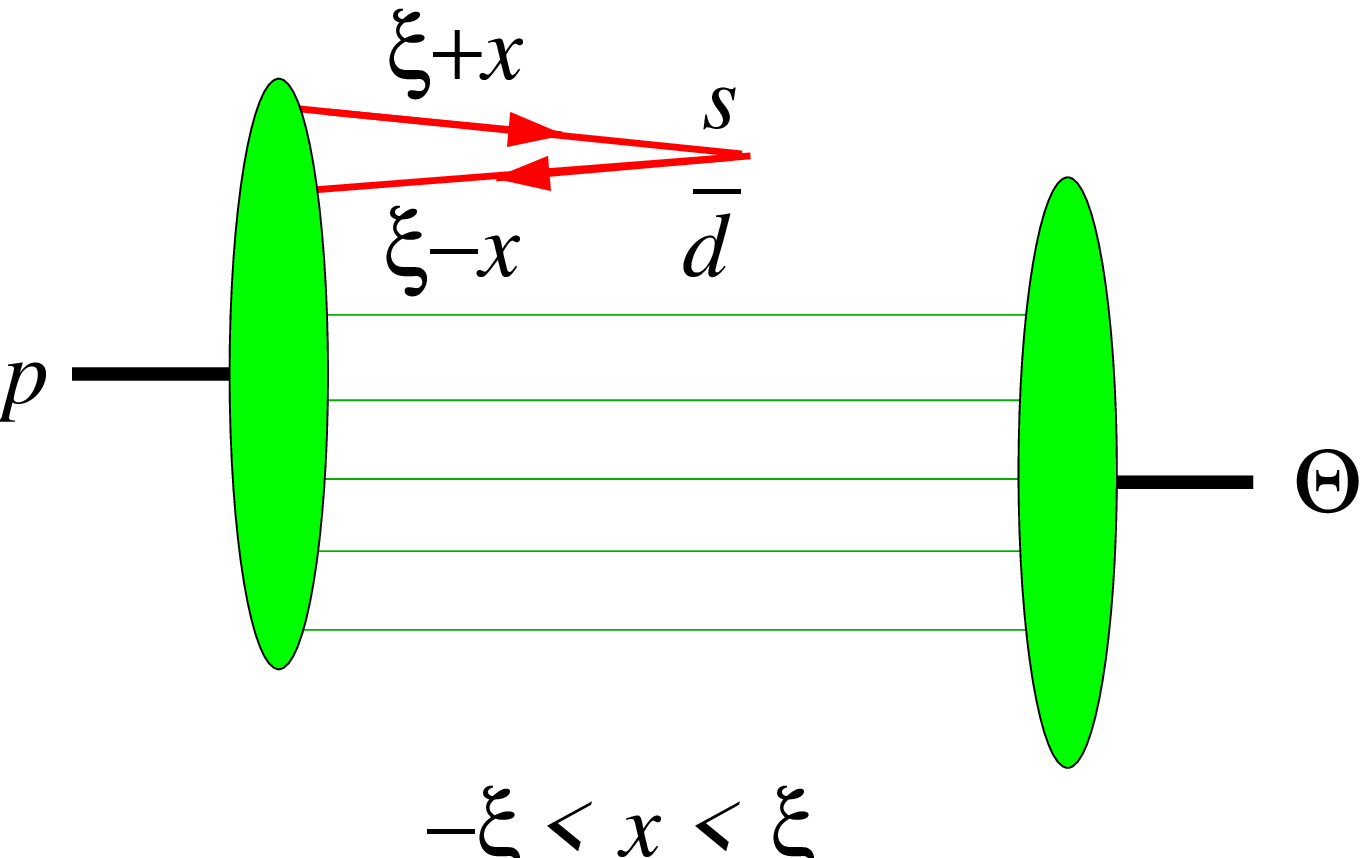,width=0.27\textwidth}
\end{center}
\caption{\label{fig:partons} Wave function representation of the
$p\to\Theta$ GPDs in the different regions of $x$.  The blobs denote
light-cone wave functions, and all possible configurations of
spectator partons have to be summed over.  The overall transverse
position of the $\Theta$ is shifted relative to the proton.}
\end{figure}

The value of $x$ determines the partonic interpretation of the GPDs.
For $\xi<x<1$ the proton emits an $s$ quark and the $\Theta$ absorbs a
$d$ quark, whereas for $-1<x<-\xi$ the proton emits a $\bar{d}$ and
the $\Theta$ absorbs an $\bar{s}$.  The region $-\xi<x<\xi$ describes
emission of an $s\bar{d}$ pair by the proton.  In all three cases sea
quark degrees of freedom in the proton are involved.  The
interpretation of GPDs becomes yet more explicit when the GPDs are
expressed as the overlap of light-cone wave functions for the proton
and the $\Theta$.  As shown in Fig.~\ref{fig:partons}, the proton must
be in \emph{at least} a five-quark configuration for $\xi<|x|<1$ and
\emph{at least} a seven-quark configuration for $-\xi<x<\xi$. 

As shown in \cite{Burkardt:2000za}, GPDs contain information about the
spatial structure of hadrons.  A Fourier transform converts their
dependence on $t$ into the distribution of quarks or antiquarks in the
plane transverse to their direction of motion in the infinite momentum
frame.  This tells us about the transverse size of the hadrons in
question.  The wave function overlap can also be formulated in this
impact parameter representation, with wave functions specifying
transverse position and plus-momentum fraction of each parton as shown on
 Fig.~\ref{fig:partons}.  We see in particular that
for $\xi<|x|<1$ the transverse positions of all partons must match in
the proton and the $\Theta$, including the quark or antiquark taking
part in the hard scattering.  For $-\xi<x<\xi$ the transverse
positions of the spectator partons in the proton must match those in
the $\Theta$, whereas the $s$ and $\bar{d}$ are extracted from the
proton at the same transverse position (within an accuracy of order
$1/Q$ set by the factorization scale of the hard scattering process).
Note that small-size quark-antiquark pairs with net strangeness are
not necessarily rare in the proton, as is shown by the rather large
kaon pole contribution to the $p\to \Lambda$ GPDs \cite{24}.  The upshot
 of our discussion is that the
$p\to \Theta$ transition GPDs probe the partonic structure of the
$\Theta$, requiring the plus-momenta and transverse positions of its
partons to match with appropriate configurations in the nucleon.  The
helicity and color structure of the parton configurations must match
as well.


\section{Scattering amplitude and cross section}
\label{sec:scatter}

The scattering amplitude for longitudinal polarization of photon and
meson at leading order in $1/Q$ and in $\alpha_s$ readily follows from
the general expressions for meson production given in
\cite{Diehl:2003ny}.  One has
\begin{eqnarray}
  \label{amp-p}
\mathcal{A}_{\gamma^* p\to \bar{K}^0\, \Theta} &=&
i e\, \frac{8\pi\alpha_s}{27}\, \frac{f_K}{Q}\, \Bigg[
I_K \int_{-1}^1 \frac{dx}{\xi-x-i\epsilon}\, 
\Big( F_A(x,\xi,t) - F_A(-x,\xi,t) \Big)
\nonumber \\
 && \hspace{4.2em} {}+
J_K \int_{-1}^1 \frac{dx}{\xi-x-i\epsilon}\,  
\Big( F_A(x,\xi,t) + F_A(-x,\xi,t) \Big)
\, \Bigg] ,
\nonumber \\
\mathcal{A}_{\gamma^* p\to \bar{K}^{*0}\, \Theta} &=&
i e\, \frac{8\pi\alpha_s}{27}\, \frac{f_{K^*}}{Q}\, \Bigg[
I_{K^*} \int_{-1}^1 \frac{dx}{\xi-x-i\epsilon}\,
\Big( F_V(x,\xi,t) - F_V(-x,\xi,t) \Big)
\nonumber \\
 && \hspace{3.8em} {}+
J_{K^*} \int_{-1}^1 \frac{dx}{\xi-x-i\epsilon}\,  
\Big( F_V(x,\xi,t) + F_V(-x,\xi,t) \Big)
\, \Bigg] ,
\end{eqnarray}
independently of the parity of the $\Theta$.  
In (\ref{amp-p}) we have integrals
\begin{equation}
  \label{DA-integrals}
I = \int_0^1 dz\, \frac{1}{z(1-z)}\, \phi(z)  ,\;\;\;
J = \int_0^1 dz\, \frac{2z-1}{z(1-z)}\, \phi(z) ,
\nonumber
\end{equation}
over the twist-two distribution amplitudes of either $\bar{K}^{0}$ or
$\bar{K}^{*0}$.  
  Note that we cannot easily guess the relative sign of the 
transition GPDs at $x$ and $-x$, since they do not become densities 
in any kinematical limit. As a consequence we cannot say 
whether the terms with $I$ or with $J$ tend to dominate in the amplitudes
(\ref{amp-p}).

To leading accuracy in $1/Q^2$ and in $\alpha_s$ the cross section for
$\gamma^* p$ for a longitudinal photon on transversely polarized
target is
\begin{eqnarray}
  \label{X-section}
\frac{d\sigma_L}{dt} &=& 
\frac{64\pi^2 \alpha_{\mathit{em}}^{\phantom{2}} \alpha_s^2}{729}\, 
\frac{f_{K^{(*)}}^2}{Q^6}\, \frac{\xi^2}{1-\xi^2}\,
( S_U + S_T\, \sin\beta ) ,
\end{eqnarray}
 where $\beta$ is the azimuthal angle between the hadronic
plane and the transverse target spin.
The cross section for an unpolarized target is simply obtained by
omitting the $\beta$-dependent term.  To have concise expressions for
$S_U$ and $S_T$ we define
\begin{eqnarray}
  \label{gpd-integrals}
\mathcal{H}(\xi,t) &=& 
I_{K^{(*)}} \int_{-1}^1 \frac{dx}{\xi-x-i\epsilon}\,
\Big( H(x,\xi,t) - H(-x,\xi,t) \Big)
\nonumber \\
&+& J_{K^{(*)}} \int_{-1}^1 \frac{dx}{\xi-x-i\epsilon}\,
\Big( H(x,\xi,t) + H(-x,\xi,t) \Big)
\end{eqnarray}
and analogous expressions $\mathcal{E}$, $\tilde\mathcal{H}$,
$\tilde\mathcal{E}$ for the other GPDs.  For $\eth = 1$ we have
\begin{eqnarray}
  \label{H-tilde-combinations}
S_U &=& (1-\xi^2) |\tilde\mathcal{H}|^2 
+ \frac{(\mth-m_N)^2 - t}{(\mth+m_N)^2}\,
\xi^2 |\tilde\mathcal{E}|^2
- \Bigg( \xi + \frac{\mth-m_N}{\mth+m_N} \Bigg)
2\xi\, \re( \tilde\mathcal{E}^* \tilde\mathcal{H} ) \, ,
\nonumber \\
S_T &=& -\sqrt{1-\xi^2}\, \frac{\sqrt{t_0-t}}{\mth+m_N} \, 
2\xi\, \im( \tilde\mathcal{E}^* \tilde\mathcal{H} )
\end{eqnarray}
for $K$ production and
\begin{eqnarray}
  \label{H-combinations}
S_U &=& (1-\xi^2) |\mathcal{H}|^2 
- \Bigg( \frac{2\xi (\mth^2 - m_N^2) + t}{(\mth+m_N)^2} 
+ \xi^2 \Bigg) |\mathcal{E}|^2
- \Bigg( \xi + \frac{\mth-m_N}{\mth+m_N} \Bigg)
2\xi\, \re( \mathcal{E}^* \mathcal{H} ) \, ,
\nonumber \\
S_T &=& \sqrt{1-\xi^2}\, \frac{\sqrt{t_0-t}}{\mth+m_N} \, 
2\im( \mathcal{E}^* \mathcal{H} )
\end{eqnarray}
for $K^*$ production.  If $\eth = -1$ then
(\ref{H-tilde-combinations}) describes $K^*$ production and
(\ref{H-combinations}) describes $K$ production.  We see that one
cannot determine the parity of the $\Theta$ from the leading twist
cross section (\ref{X-section}) without knowledge about the dependence
of $\mathcal{H}$, $\mathcal{E}$, $\tilde\mathcal{H}$,
$\tilde\mathcal{E}$ on $t$ or $\xi$.  

There are arguments 
that theoretical uncertainties 
 cancel at least partially in suitable ratios of cross
sections.  Processes to compare with are given
by $ep \to e K^0 \Sigma^+$, $ep \to e K^+ \Sigma^0$, $ep \to e K^+
\Lambda$ or their analogs for vector kaons, with
the production of either ground state or excited hyperons.  
Their amplitudes are given as in
(\ref{amp-p}) with an appropriate replacement of matrix elements $F_V$
or $F_A$ listed in Table~\ref{tab:channels}.  
Isospin invariance further gives $F_{p\to \Sigma^+}
= \sqrt{2}\, F_{p\to \Sigma^0}$.

\begin{table}
\caption{\label{tab:channels} Combinations of transition GPDs
multiplying $I$ and $J$ in the hard scattering formula
(\protect\ref{amp-p}) and its analogs for the listed channels.}
$$
\renewcommand{\arraystretch}{1.2}
\begin{array}{lll} \hline\hline
  & ~~~~~~~~~~~~~~~I  & ~~~~~~~~~~~~~~~J  \\ \hline
\gamma^* p \to \bar{K}^0 \Theta &  
\phantom{-} F_{p\to \Theta}(x) - F_{p\to \Theta}(-x) &
\phantom{-} F_{p\to \Theta}(x) + F_{p\to \Theta}(-x) \\
\gamma^* p \to K^0 \Sigma^+ &
\phantom{-} F_{p\to \Sigma^+}(x) - F_{p\to \Sigma^+}(-x) &
- [ F_{p\to \Sigma^+}(x) + F_{p\to \Sigma^+}(-x) ] \\
\gamma^* p \to K^+ \Sigma^0 &
- [ 2F_{p\to \Sigma^0}(x) + F_{p\to \Sigma^0}(-x) ] &
\phantom{-} 2F_{p\to \Sigma^0}(x) - F_{p\to \Sigma^0}(-x) \\
\gamma^* p \to K^+ \Lambda &
- [ 2F_{p\to \Lambda}(x) + F_{p\to \Lambda}(-x) ] &
\phantom{-} 2F_{p\to \Lambda}(x) - F_{p\to \Lambda}(-x) \\ \hline
\end{array}
\renewcommand{\arraystretch}{1}
$$
\end{table}

\section{Conclusions}
\label{sec:concl}

We have investigated exclusive electroproduction of a $\Theta^+$
pentaquark on the nucleon at large $Q^2$, large $W^2$ and small $t$.
Such a process provides a rather clean environment to study the
structure of pentaquark at parton level, in the form of well defined
hadronic matrix elements of quark vector or axial vector currents.  In
parton language, these matrix elements describe how well parton
configurations in the $\Theta$ match with appropriate configurations
in the nucleon (see Fig.~\ref{fig:partons}).  Their dependence on $t$
gives information about the size of the pentaquark.  Channels with
production of pseudoscalar or vector kaons and with a proton or
neutron target carry complementary information.  The transition to the
$\Theta$ requires sea quark degrees of freedom in the nucleon, and we
hope that theoretical approaches including such degrees of freedom
will be able to evaluate the matrix elements given in
(\ref{matrix-elements}).  

At modest values of $Q^2$ the leading
approximation in powers of $1/Q^2$ and of $\alpha_s$ on which we based
our analysis may receive considerable corrections.  The associated
theoretical uncertainties should be alleviated by comparing $\Theta$
production to the production of $\Sigma$ or $\Lambda$ hyperons as
reference channels.  In any case, even a qualitative picture of the
overall magnitude and relative size of the different hadronic matrix
elements accessible in the processes we propose would give information
about the structure of exotic baryons (see also Ref. \cite{APSTW} 
for the case of hybrid mesons) well beyond the little we presently
know. 


\section*{Acknowledgments}  

The work of B. P. and L. Sz.\ is
partially supported by the French-Polish scientific agreement
Polonium.  CPHT is Unit{\'e} mixte C7644 du CNRS.


\end{document}